\documentclass[useAMS,usenatbib]{mn2e}
\usepackage{psfig}

\def\ltsima{$\; \buildrel < \over \sim \;$}
\def\lsim{\lower.5ex\hbox{\ltsima}}
\def\gtsima{$\; \buildrel > \over \sim \;$}
\def\gsim{\lower.5ex\hbox{\gtsima}}
\def\be{\begin{equation}}
\def\ee{\end{equation}}
\def\mec2{m_{\rm e}c^{2}}

\def\msec{{\rm M_{\sun} sec^{-1}}}
\def\ros{r_{\rm d_7}}
\def\ro{r_{\rm d}}
\def\m{m_{\rm fb}}
\def\no{\noindent}

\newcommand{\mnras}{MNRAS }

\newcommand{\aap}{A\&A }
\newcommand{\apj}{ApJ }

\newcommand{\prd}{Phys.~Rev.~D }

\begin{document}
\title[Observing nearby discs]{Delayed X-ray emission from fallback in compact-object mergers}

\author[Rossi \& Begelman] {Elena M. Rossi$^{1,2}$ \&
Mitchell C. Begelman $^{1,3}$\\
$^1$JILA, University of Colorado at Boulder, 440 UCB, Boulder, CO 80309-0440 \\
$^2$Chandra Fellow \\
$^3$Department of Astrophysical and Planetary Sciences, University of Colorado \\
\tt e-mail: emr@jilau1.colorado.edu; mitch@jila.colorado.edu}

\maketitle
  
\begin{abstract}
When double neutron star or neutron star-black hole binaries merge, the
final remnant may comprise a central solar-mass black hole surrounded
by a $\sim 0.01-0.1~ M_{\sun}$ torus. The subsequent evolution of this
disc may be responsible for short $\gamma$-ray bursts (SGRBs).  A
comparable amount of mass is ejected into eccentric orbits and will
eventually fall back to the merger site after $\sim 0.01$ seconds.  In
this {\it Letter}, we investigate analytically the fate of the
fallback matter, which may provide a luminous signal
long after the disc is exhausted.  We find that matter in the
eccentric tail returns at a super-Eddington rate and is eventually
($\gsim 0.1$ sec) unable to cool via neutrino emission and accrete all the
way to the black hole.  Therefore, contrary to previous claims, our
analysis suggests that fallback matter is {\it not} an efficient
source of late time accretion power and is unlikely to cause the late
flaring activity observed in SGRB afterglows.  The fallback matter
rather forms a radiation-driven wind or a bound atmosphere.
In both cases, the emitting plasma is very opaque and photons are released with
a degraded energy in the X-ray band. We therefore suggest that
compact binary mergers could be followed by an ``X-ray renaissance",
as late as several days to weeks after the merger.
This might be observed by the next generation of X-ray detectors.
\end{abstract}

 
\section{Introduction}

Close binaries of compact solar-mass objects are expected to form via
the evolution of massive star binaries or by dynamical interaction in
dense star clusters.  Neutron star (NS--NS) binaries have been
detected as radio pulsars \citep[e.g.][]{faulkner05}, and while
black hole--neutron star (BH-NS) or double black hole (BH-BH)
binaries have not been observed directly, they are predicted by
population synthesis models.  The compact objects are expected to
merge due to gravitational wave emission, with evolutionary scenarios
estimating a local rate of NS--NS mergers $10-100$ times higher than
for BH-NS and BH-BH systems \citep[e.g.][]{bel07}. The final remnant
for NS--NS and NS--BH coalescence is generally thought to be a BH of a
few solar masses surrounded by a $0.01-0.1~ M_{\sun}$ accreting disc
\citep[e.g.][]{ruffert97,shibata03,ross04,faber06}.  The accretion
power immediately following the merger is perhaps the ultimate cause
of SGRBs \citep{blinnikov84,eichler89,paczynski91}.  At early times
($\lsim 0.1 -1$ sec), the accreting disc is geometrically thin,
effectively cooled by neutrino emission \citep{popham99}. When the
accretion rate drops below $\sim 0.1 ~\msec$ --- the exact value
depending on the accretion parameter $\alpha$ and BH spin
(Chen \& Beloborodov 2007; Metzger, Piro \&  Quataert 2008) --- the disc becomes radiatively inefficient
and super-Eddington accretion drives a substantial outflow
\citep{metzger08}.

During the dynamical phase of the merger, in which the lighter
companion is tidally disrupted, a fraction ($\sim 10^{-2} M_{\sun}$)
of the debris receives enough energy to be ejected from the system
while a comparable amount remains bound in eccentric orbits
\citep[e.g.,][]{ross07,faber06} and will eventually return to the disc
site: {\em fallback} matter.  This weakly bound matter may give rise
to interesting phenomena observable on timescales longer than any
viscous timescale of the disc.  For example, it has been suggested
\citep{lee07,ross07,metzger08} that it can be responsible for the X-ray
flaring, observed in SGRB afterglows on timescales of minutes to hours
\citep[e.g.][]{campana06}.  Unfortunately, numerical investigations
have not yet been able to follow the long-term ($>$ minutes) evolution
of this eccentric tail, because of time-step limitations
\citep{ross07}.

In this {\em Letter}, we investigate analytically the fate of matter
falling back onto a recent merger.  We argue that energy released
during fallback is {\it not} a promising source of the X-ray
flares. The energy liberated during fallback will either lead to a
powerful, radiation-driven wind or a more gradually expanding ``breeze''
that could ultimately form a bound cloud around the merged object.  In
either case, the expanding gas is so opaque that the radiation is
trapped in the expanding flow and degraded to low energies before
being released in the X-ray band.  We therefore suggest that
compact binary mergers might be accompanied by delayed X-ray
emission.  We assess the detectability of this emission 
when the merger is localized by
either a short $\gamma$-ray burst or a gravitational wave signal.
 A
direct observation of the accretion activity would give us valuable
information on how compact-object binaries merge.

This {\it Letter} is organized as follows. We discuss the
behavior of the fallback matter in \S~\ref{sec:fb}.
Then, we consider two possible scenarios for this material as it rebounds: we model
a wind in \S~\ref{sec:wind} and a bound atmosphere in \S~\ref{sec:atmo}.
Prospects for detecting the X-ray emission are discussed in \S~\ref{sec:detection} and
conclusions are drawn in \S~\ref{sec:conclu}.


\section{accretion behavior of fallback matter}
\label{sec:fb}
 In our analysis, we scale our parameters with values appropiate for
 NS-NS binaries, since these systems are the most common.  The
 encounter of a couple of neutron stars is followed by the formation
 of a central attractor with typical mass $M_{\rm c}= 2.5~m_{\rm
 c}~M_{\sun}$ \citep{bel08}, surrounded by an accretion disc that
 extends initially up to $r_{\rm d}= 10^7 r_{\rm d_7}~ {\rm cm}$
 \citep[e.g.,][]{ruffert97}.  The weakly bound material, $M_{\rm fb}=
 3 \times 10^{-2} \m~M_{\sun}$ \citep{ross07}, launched into elliptical orbits, will
 travel as far as its apocenter and eventually come back to its
 pericenter $r_{\rm p} \simeq r_{\rm d}$.  The
 rate at which this material accretes can be found analytically
 assuming that the energy distribution with mass is flat: the
 accretion rate, after a plateau phase at $\dot{M}_{\rm max}$,
 decreases with time as
 
 \be
 \dot{M}_{\rm fb}(t) = \dot{M}_{\rm max} \left(\frac{t_{\rm min}}{t}\right)^{5/3},
 \label{eq:mdotfb}
 \ee

\no
 \citep{phinney89}, where the minimum arrival time
corresponds to the period of orbits with eccentricity $e\simeq 0$,

\be
t_{\rm min} \simeq \frac{2\,\pi \, \ro^{3/2}}{\sqrt{G M_{\rm c}}} \simeq 10^{-2}~ m_{\rm c}^{-1/2} r_{\rm d_7}^{3/2}~{\rm sec},
\ee

\no
and the initial accretion occurs at a rate 
\be
\dot{M}_{\rm max} = \frac{2}{3}\; \frac{M_{\rm fb}}{t_{\rm min}} \simeq 2~ \frac{\m~m_{\rm c}^{1/2}}{r_{\rm d_7}^{3/2}}~ \msec.
\label{eq:dotmax}
\ee

\no
 Even if eq.~\ref{eq:mdotfb} has been originally derived for tidal distruption of stars in the potential well of a supermassive black hole, numerical calculations by \cite{ross07} indicate that the $t^{-5/3}$ law also applies to the case we are investigating.

When the fallback matter hits the disc (or the leftover material),
 its kinetic energy (per unit mass) $v_{\rm fb}^2/2 \simeq GM_{\rm c}(-1/(2
 a)+1/\ro) \sim G M_{\rm c}/\ro$ is converted into heat via shocks. 
The internal energy of the shocked matter is photon-dominated.
Initially, the fallback matter would simply join the disc and accrete
onto the central object, because it is effectively cooled by neutrino
emission: i.e., the flow rate is {\it sub}-Eddington with respect to
the neutrino luminosity and accretion is possible. The neutrino emissivity $q_{\nu}^- = q_{\rm an}+q_{\rm eN}$
is due both to electron-positron pair annihilation 
$q_{\rm an} \propto T_{\rm sh}^9$ and capture onto nuclei $q_{\rm eN} \propto T_{\rm sh}^6 \rho_{\rm fb}$
\citep[see][for the analytic approximations]{popham99}.
The fallback matter density at $\ro$  is $\rho_{\rm fb} = \dot{M}_{\rm fb}/(4 \pi \ro^2 v_{\rm fb})$,
while the temperature to which the gas is shock-heated can be approximately obtained
by equating its kinetic energy  density  at $\ro$, ($v_{\rm fb}^2\rho_{\rm bf}/2$), to its internal energy density,
\be
T_{\rm sh} = \left(\frac{G\,M_{\rm c}\,\rho_{\rm fb}}{\ro\,a_{\rm r}}\right)^{1/4} = 3.6 \times 10^{10} \left(\frac{t_{\rm min}}{t}\right)^{5/12}{\rm K}, 
\label{eq:tsh}
\ee

\no
where $a_{\rm r}$ is the radiation constant.
The BH feeding happens for
large enough accretion rates, $\dot{M} > \dot{M}_{\rm ign} \simeq 0.14~\msec$,
when the cooling time $t_{\rm c} = a_{\rm r}T_{\rm sh}^{4}/q_{\nu}^-$  is shorter than the viscous time at $\ro$
$t_{\rm vis} \simeq 0.18~\alpha_{0.1}^{-1} m_{\rm c}^{-1/2}\ros^{3/2}\left(H/(\ro0.3)\right)^{-2}$ sec.
When $t=t_{\rm w} \simeq 5 \times 10^{-2} $ sec, the accretion rate $\dot{M}_{\rm fb}$ drops below the critical value
$\dot{M}_{\rm ign}$ and neutrino cooling becomes inefficient and eventually (when $k_{\rm b}T_{\rm sh}< \mec2$) switches off completely.
The remaining reservoir of mass in the tail is still
substantial $M_* = (3/2)~ \dot{M}_{\rm bf}(t_{\rm w})~ t_{\rm w}\simeq 7 \times 10^{-3} M_{\sun}$
and, unable to accrete onto the black hole, it is likely to be blown off the disc plane.


\section{wind model}
\label{sec:wind}
The first possible fate for the fallback matter that we consider is
the formation of a radiation-driven wind.  

The amount of mass
entrained in the wind $M_{\rm w}$ and
 its specific energy are uncertain. 
One possibility is that the total kinetic energy of fallback matter
is deposited unevenly, so that a small fraction of mass ($M_{\rm w}\ll M_*$)
can reach a final velocity that exceeds the escape velocity and form a wind. On the other 
extreme, the wind could have an amount of mass {\em comparable} to the fallback tail
$M_{\rm w}\simeq M_*$,  where sufficient internal energy to unbind this weakly bound matter is
gained  via accretion.
 Given the range of uncertainty,  we will scale our equations adopting
$M_{\rm w}= 10^{-3} M_{\rm w_{-3}} M_{\sun}$ and a terminal velocity
$v_{\rm t} = 0.3c ~\beta_{0.3}$, where c is the speed of light and we
use as guidance the escape velocity $v_{\rm esc}=\sqrt{2G M_{\rm
c}/r_{\rm d}}= 0.27 c$ at $r=r_{\rm d}$.

To model the wind, we take an initial radius $r_{\rm 0} \simeq r_{\rm
d} $. This is sufficiently close to the sonic radius that we may
assume, in first approximation, an outflow with constant velocity
equal to its terminal velocity $v_{\rm t}$\footnote{For a polytropic wind
with $\gamma=4/3$ the velocity at the sonic point is only $\sqrt{3}$
smaller than the terminal velocity.}.  The wind, powered by fallback
matter, will steadily decrease with time according to
eq.~\ref{eq:mdotfb},
$  \dot{M}_{\rm w}(t) =
3.4 \times 10^{23}~ M_{\rm w_{-3}} t_{\rm w_{-1}}^{2/3}~t_{\rm hr}^{-5/3}~ {\rm g\, sec^{-1}},
$
where  $t_{\rm w}=  0.1~t_{\rm w_{-1}}$ sec and the time $t$ since the onset of the wind is in hours ($t_{\rm hr}$).
Its matter density then follows from matter conservation,
\be
\rho(r,t) = \frac{\dot{M}_{\rm w}(t)}{4 \pi r^2 v_{\rm t}},
\ee

\no 
for radii $r< v_{\rm t}t$.
The radiation pressure $P= (1/3)~ a_{\rm r} T^{4}$  can be related to $\rho$ by the polytropic
relation, with index $4/3$. Therefore, the temperature decreases slowly as
\be
T(r,t) \propto P^{1/4} \propto r^{-2/3} t^{-5/12}.
\label{eq:Trt}
\ee

 The radiation transported with the wind is mostly liberated at the
 trapping radius $r_{\rm tr}$ where the diffusion timescale for
 photons equals the expansion timescale.
 Beyond this radius, the luminosity is transported by radiative flux
 up to the photosphere, where the optical depth $\tau \sim 1$.
 In our case $\beta \sim 0.3$ or higher,
 therefore the trapping radius is very close to the photospheric radius
  and we will ignore in the following the radiative layer.
 The optical depth for electron scattering $\tau
 \simeq \rho \kappa r $ is computed with a Thomson opacity
 $\kappa=0.2~ \kappa_{0.2}$, that we scale with the value appropriate
 for a flow composed solely of $\alpha$-particles.  The electron
 density is, in fact, uncertain: it depends mainly on the initial
 composition of the wind at $\ro$\footnote{In principle the
 neutrino/antineutrino luminosities from the disc can change the 
proton-tp- neutron ratio in the flow. However, at timescales of
 interest to us, the neutrino emission has died off.}, which includes
 $\alpha$-particles and free baryons. For a neutron-rich composition,
 $\kappa < 0.2 $. The nucleosynthesis in the wind does not change the
 free electron density, since temperatures are high enough for the
 recombined helium to be fully ionized.
 The trapping radius then reads :

\be
r_{\rm tr} = \frac{\dot{M}_{\rm w} \kappa}{4\pi c} 
 \simeq  1.8 \times 10^{11}~\kappa_{0.2}~ M_{\rm w_{-3}}~ t_{\rm w_{-1}}^{2/3}~ t_{\rm hr}^{-5/3} ~{\rm cm}.
\label{eq:rtr}
\ee

\no
Conservation of energy $1/2 \dot{M}_{\rm w} v_{\rm t}^2  \approx 16 \pi (a/3)T_{0}^4 \ro^2 v_{\rm t}$  
allows us to solve for the central temperature,
$T_{0}(\ro,t) \simeq 10^{8} ~\beta_{0.3}^{1/4}~ M_{\rm w_{-3}}^{1/4}~t_{\rm w_{-1}}^{1/6}/ \ros^{1/2}  t_{\rm hr}^{-5/12} ~{\rm K},$
and from eq.~\ref{eq:Trt} we can derive the temperature at the trapping radius

\be
 T_{\rm tr}(r_{\rm tr},t)\simeq 1.5 \times 10^{5} \left(\frac{\beta_{0.3}^{1/4}\ros^{1/6}} {\kappa_{0.2}^{2/3}~M_{\rm w_{-3}}^{5/12}~t_{\rm w_{-1}}^{5/18} } \right)
~ t_{\rm hr}^{25/36} ~{\rm K}.
\label{eq:txwind}
\ee

\no
The emission from the trapping radius of the wind becomes harder with time while the luminosity, 
$L_{\rm tr}(r_{\rm tr},t) = \frac{16 \pi}{3} ~aT_{\rm tr}^4~ v_{\rm t}\, r_{\rm tr}^2$, decreases,

\be            
L_{\rm tr}(r_{\rm tr},t) \simeq 2 \times 10^{40} \beta^2_{0.3}\,\ros^{2/3} \kappa_{0.3}^{-2/3}
M_{\rm w_{-3}}^{1/3}~t_{\rm w_{-1}}^{2/9}~ t_{\rm hr}^{-5/9} ~{\rm erg\,sec^{-1}}.
\label{eq:lxwind}
\ee

\no
When $r_{\rm tr} = r_{\rm d}$,

\be
t_{\rm x} \simeq  14.9 ~\left(\frac{\kappa_{0.2}~M_{\rm w_{-3}}}{\ros}\right)^{3/5}~t_{\rm w_{-1}}^{2/5} ~{\rm days},
\label{eq:tx}
\ee

\no
the thermal emission has a temperature of

\be
T_{\rm x}(\ro,t_{\rm x}) \simeq 9 \times 10^{6}  \left(\frac{\beta_{0.3}}{\ros \kappa_{0.2}}\right)^{1/4}~{\rm K},
\label{eq::Tx}
\ee
 
 \no
 and a luminosity
 
 \be
 L_{\rm x} (\ro,t_{\rm x}) \simeq  7.5 \times 10^{38} \beta_{0.3}^2 \ros \kappa_{0.2}^{-1} ~{\rm erg\,sec^{-1}}.
\label{eq:lx}
\ee

\no 
 We note that in eq.~\ref{eq::Tx} and eq.~\ref{eq:lx} the only
 dependences are on the initial radius, the terminal velocity and the opacity.  The
 dependence is particularly weak for $T_{\rm x}$ because,
 when $r_{\rm tr} = r_{\rm d}$, the accretion rate is set only by the
 size of the launching region and by the opacity,
(see eq.~\ref{eq:rtr}). 
Therefore
$L_{\rm x} \simeq \dot{M}_{\rm w}v_{\rm t}^2 \propto \ro\,v_{\rm t}^2/ \kappa   \propto \ro^{2} T_{0}^4  v_{\rm t}$.  
We stress the important role of the wind composition, equivalently of the electron fraction.
For an extreme proton-to-neutron ratio of $0.1$, $\kappa \simeq 0.04$ and the X-ray 
emission ($T_{\rm tr}\geq 10^{6}$ K) starts at $t_{\rm x} \simeq 3 ~{\rm hr}$ with a  luminosity  
$L_{\rm tr} \simeq 3.4 \times 10^{40}$ erg/sec.

The emission from the wind will switch off when the whole energy
supplied by the wind can be accreted ($\dot{M}_{\rm fb} \simeq M_{\rm
edd} \simeq 7 \times 10^{18}{\rm g/sec} $).  This is a long time of the order of $\sim 3$ months.
However, when $L_{\rm x}$ drops below $\sim L_{\rm x}/2$,
the X-ray emission is likely to be
undetectable.  This happens for $t / t_{\rm x} \gsim $ a few.

\section{Atmosphere model}
\label{sec:atmo}

 Another scenario may be envisaged where a bound atmosphere forms
 around the central object.  This can happen, if the outflowing gas
 retains the same amount of energy per unit mass that the eccentric
 tail had. This gas would still be bound to the central BH: it would
 start expanding from $\ro$ nearly isotropically until it reaches a
 radius $r_*$, where its internal energy is $\sim$ half its potential
 energy. After a few seconds, this inflated gas cloud has a nearly
 constant mass $M_* \sim \dot{M}_{\rm fb}(t_{\rm w})\times t_{\rm w}
 \simeq 4.6 \times 10^{-3} {\rm M_{\sun}}$ and radius $r_*$, since
 most of the mass $M_{\rm *}$ is injected around $t\sim t_{\rm w}$.
 We can estimate the radius $r_*$ through $GM_{\rm c} M_*/(2r_*)
 =\int_{\ro}^{\infty} GM_{\rm c}/(2a) (dm/da)\, da$, where $a$ is the
 semi-major axis of the particle orbits in the eccentric tail.  Since
 the distribution of specific orbital energy $\epsilon = -GM_c/(2a)$
 with mass is constant $dm/d\epsilon \simeq M_*/\Delta \epsilon$,
 where $\Delta \epsilon \sim GM_{\rm c}/\ro$ is the extra energy gained by $M_*$ via the
 tidal torque, we can solve the integral and find $r_*/\ro \simeq
 \Delta \epsilon / (GM_{\rm c}/(2\ro))$. We conclude
 that $r_*$ is of the same order as $\ro$. This reflects the fact 
 that most of the mass is at small $a$.
 We choose to parametrize $r_* = 10~ \ro = 10^8~ r_{*8}~ {\rm cm}$.
 
 We can then calculate the cloud's mean properties. Its mean density is
 $\rho_* = 2.2 \times 10^6 r_{*8}^{-3}~{\rm {\rm g}~cm^{-3}}$ and the
 temperature can be derived equating its internal energy density (in
 radiation and gas) with $GM_{\rm c}\rho_*/(2r_*)$.
 Radiation pressure is $\sim 5$ times higher than gas pressure and
 temperature decreases linearly with the cloud radius, $T \simeq 4.6
 \times 10^{9} r_{*8}^{-1}~{\rm K}$ (while the pressure ratio remains
 constant).  The gas cloud is in hydrostatic equilibrium, since it changes
 its properties on a time scale $M_*/\dot{M_*} \simeq 4 \times 10^6~
 {\rm sec}~ t_{\rm hr}^{5/3}$, much longer than the dynamical timescale
 $t_{\rm dy} \simeq 2.6~ {\rm sec}~ r_{*8}^{3/2}$. 
 The rotational energy may be neglected, being a factor $\sim
 (\ro/r_*)$ times smaller than the internal energy.
 On timescales of interest, the cloud does not deflate via radiative losses,
since the diffusion time is very long, $t_{\rm diff} = r_*^2 \rho_* \kappa/c
 \sim 4600~ r_{*8}^{-1}~ {\rm yrs}$.
 The merged object is thus surrounded and 
 obscured by a persistent source, emitting at the Eddington limit 

\be
L_{\rm ph} = L_{\rm edd}= \frac{4 \pi GM_{\rm c}\,c}{\kappa} = 6.2 \times 10^{38}  \kappa_{0.2}^{-1}~ {\rm erg~ sec^{-1}},
\label{eq:ledd}
\ee

\no
at a temperature $T_{\rm ph} \simeq 3.1 \times 10^{6} r_{*8}^{-1/2} {\rm K}$.

\section{Detection prospects}
\label{sec:detection}
 \subsection{X-ray signal}
\label{sec:xray}

After a few minutes, the photons escaping from the wind are in the
ultraviolet band and after an hour or so in the extreme ultraviolet
 (EUV). This emission is strongly absorbed and unlikely to be
observed. At later times $t\simeq t_{\rm x}$,
however, the emission should peak in soft X-ray, with a luminosity
$\sim L_{\rm edd}$ (eqs.~\ref{eq:lx} and~\ref{eq:ledd}) and a thermal spectrum with temperature $\sim 0.8~
{\rm keV}$ (eq.~\ref{eq::Tx}). In the case that an opaque cloud surrounds the
merged object, we have comparable luminosity, emitted at a temperature 
that depends on the extension or the atmosphere: for simplicity, 
we consider here the case in which 
the emission is at $\sim 0.8~{\rm keV}$, corresponding to $a \sim 5 \times 10^{7}$ cm.
The main difference with the wind case is that this emission should be persistent.

 Under favorable environmental conditions, 
  the emission may be observable. 
 If the merger occurs in a galactic
 halo or even in the intergalactic medium, absorption should be
 moderate. Moreover, those locations may not be polluted by
 contaminating soft X-ray sources. Finally, compact-object mergers
 should not be accompanied by a bright supernova explosion,
 eliminating another possible co-located X-ray source.
 
An Eddington luminosity yields an unabsorbed flux at redshift $z$
 of $F = 3.3 \times
 10^{-16} \left(0.03/z\right)^2 ~{\rm
 erg\, cm^{-2} sec^{-1}}, $ where we have approximated the
 luminosity distance at redshift $z$ as $D_{\rm l}(z) = ~4.2 \times
 10^{3} (H_{o}/71)^{-1}~z~ {\rm Mpc}$.  Simulating the response of
 different current and future instruments
\footnote{http://heasarc.gsfc.nasa.gov/Tools/w3pimms.html},  allows us to determine the expected count rate as a function 
of redshift. Assuming $N_{H} = 10^{20} {\rm cm^{-2}}$ (Galactic and intrinsic to the host galaxy) and a black body spectrum,
we get a count rate
$\phi = K_{\rm in} \left(\frac{0.03}{z}\right)^2~ {\rm cts\,sec^{-1}},$ where $K_{\rm in} \simeq 5.4 \times 10^{-5} $ for
{\it XMM}, $K_{\rm in} \simeq 4.6 \times10^{-5}$ for {\it Chandra},
$K_{\rm in} \simeq 4.5 \times 10^{-3}$ for {\it XEUS} and $K_{\rm in}
\simeq 1.1 \times 10^{-3}$ for {\it Con-X}.
 
Fig.~1 shows that  X-ray detection is most likely to be feasible 
with the next generation of instruments. 
 The proposed missions 
{\it Con-X} and {\it XEUS} will be able to 
collect $\gsim 10$ cts or more in a $10^{5}$ sec exposure from mergers occurring 
as far as $z \simeq 0.1$ and $z \simeq 0.2$ respectively.
The local merger rate of NS-NS is estimated to be $\sim 0.8-10~ \times
10^{-5} {\rm yr^{-1}}$ per Milky Way galaxy
\citep{bel07,kim06}\footnote{The quoted numbers are the minimum and
maximum theoretically estimated numbers in \citet{bel07}.  The main
source of uncertainty is the treatment of the ``common envelope"
channel to compact-object formation.}.  If half of those
systems eventually merge in the halo, their rate is $\sim 0.4-5~
\times 10^{-7} {\rm mergers~ yr^{-1} Mpc^{-3}}$, where a number
density of $0.01~{\rm galaxies~Mpc^{-3}}$ as been assumed
\citep{oshau08}.
 Combining the NS-NS merger rate and the instrument observable volumes,
 we predict that {\it Con-X} could observe $\sim 13-156  ~{\rm
 mergers ~yr^{-1}}$ while the expected rate for ${\it XEUS}$
 is $\sim 100-1251~ {\rm mergers ~yr^{-1}}$.
\subsection{Merger localization}

In order to detect the X-ray emission, it is necessary to localize the merger. In the following, we
discuss two possibilities.
\subsubsection{Short GRBs} 
\label{sub:grbs}

Coalescence of compact objects (especially NS-NS) are possible candidates as progenitors of
SGRBs \citep[see][for a review]{nakar07}.
Therefore, in principle, a binary coalescence can be localized via a short burst, though
there may be limitations.
First, the local observed rate is estimated
to be $\sim 100$ times smaller than the local rate of compact object
mergers. The discrepancy is mostly credited to the geometrical
beaming of the burst jet.  
Moreover, the {\it observed} redshifts range typically between $0.1- 1.5$.
However, the distances of these
sources could only be measured for a handful of cases in the
last few years. Despite this, it is reasonable to assume that short GRBs should also
explode closer to us, if they are indeed produced by NS-NS mergers, and that
some selection effects are preventing us from measuring their
redshifts. When a short GRBs with $z \lsim 0.2$ should be localized, 
the X-ray emission from the wind or the atmosphere could be brighter than the X-ray afterglow 
around two weeks later.

\subsubsection{Gravitational waves as signposts} 
\label{sub:gw}

The gravitational wave signal is a more promising signpost for
mergers.
This is because this signal should be associated 
with any coalescence of compact objects, 
unlike the beamed $\gamma$-ray emission from SGRBs.
An instrument such as {\it advanced LIGO}\footnote{http://www.ligo.caltech.edu/advLIGO/}
 should be able to detect
mergers of two neutron stars to a distance of $z \sim 0.07$: this
implies a detection rate for {\it XEUS} and {\it Con-X} of $4-54 ~{\rm
mergers ~yr^{-1}}$.  Advanced {\it LIGO} will be, in fact, more
sensitive to NS-BH binaries, which should be visible up to $z\simeq
0.15$.  Since the Galactic merger rate for these systems is $0.1-5
\times 10^{-6}~{\rm yr^{-1}}$ per galaxy \citep{bel07}, {\it XEUS} is
expected to detect $1-53$ such mergers per year, while {\it Con-X}
less than $15$ per year.

The main limitation seems to be how accurately the merger position
could be localized.  Current estimates suggest that a network of
non-collocated advanced interferometers --- such as advanced {\it LIGO},
advanced {\it
VIRGO}\footnote{http://wwwcascina.virgo.infn.it/advirgo/} and {\it LCGT}, \citet{kazuaki06})---will be
able to detect inspiraling binaries at the redshifts of interest and
localize them at a degree level \citep{syl03}. This is enough for an
optical but not for an X-ray follow-up. However, {\bf with a solid
angle error} ten times smaller, we can identify a region of
the sky with only one local galaxy with redshift $z\lsim 0.05$;
for a galaxy at $z\lsim 0.15$, the localization error 
should be, instead, a few hundreds times smaller. 
The source distance maybe be obtained directly from the
gravitational signal (Abbot at al. 2008).  This would greatly help the
search for the X-ray fallback signal.

\begin{figure}
\psfig{figure=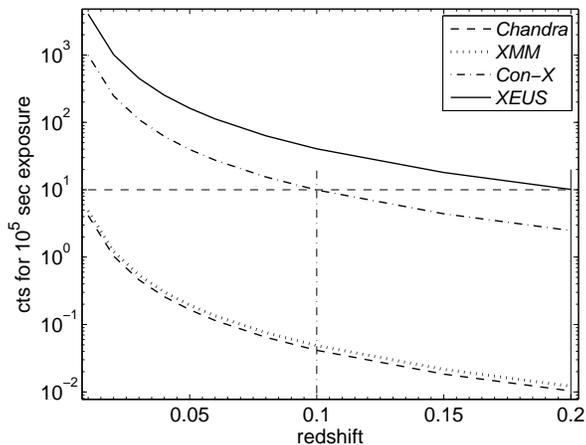,width=0.48\textwidth}
\caption[]{Counts for an exposure of $10^{5}$ sec for the X-ray
telescopes quoted in the legend. The maximum distance for which they
collect more than 10 couts (horizontal solid line) is indicated by the
corresponding vertical lines.}
\label{fig:cts1d5}
\end{figure} 

\section{Discussion and Conclusions}
\label{sec:conclu}

In this {\it Letter}, we investigate the possible fate
of fallback matter associated with mergers of compact objects, where a disc is formed by
disruption of a NS.  Matter flung to highly eccentric orbits will
eventually come back to the disc at a super-Eddington rate, converting
its kinetic energy into heat via shocks, and will be unable to cool by
neutrino emission. Contrary to previous claims, 
we think that this implies that fallback matter {\it cannot} accrete all the way to the central object and be 
responsible for the late energy injections observed in GRB afterglows.
Rather, the fallback matter is likely to be blown off the disc plane, leading to
the formation of a radiation-driven wind or a bound atmosphere.
For the wind case, we have analytically calculated the time evolution of the temperature
and luminosity at the trapping radius: while the luminosity decreases (eq.~\ref{eq:lxwind}),
the wind photosphere becomes hotter (eq.~\ref{eq:txwind}).  At first, the emission is in the
EUV band and absorption will likely prevent us from observing
it. After one or two weeks, the emission finally peaks in the soft
X-ray band and the wind activity can be observed. The bound cloud is radiation pressure-dominated and
emits at the Eddington limit in soft X-rays, if the atmosphere 
does not extend much further than $10^{8}$ cm.
We note that our estimates of luminosities are conservative:
factors such as a smaller electron fraction in the ejected plasma
and moderate geometrical beaming can substantially increase 
the expected luminosity.

 We also discuss detection prospects for this delayed X-ray activity.
Our inspection indicates that only in
fortuitous circumstances could the X-ray emission be detected with
current instruments, while the planned missions (such as Con-X and XEUS) have a better
chance \S~5.1. Then, the main limiting factors will not be the X-ray
detector capability, but rather the tool for localizing the merger \S~5.2. On the
one hand, short $\gamma$-ray bursts can be easily detected and localized 
in the whole volume where instruments like {\it Con-X} and {\it XEUS} can observe
the X-ray emission; however, they are estimated to occur at a rate that is
$\sim 100$ times smaller than the rate at which compact binaries merge.
On the other hand, the planned advanced gravitational wave
interferometers should be able to detect a signal from
any such a merger but within cosmic distances smaller than the maximum
distance that {\it Con-X} and {\it XEUS} can reach. Moreover, X-ray follow up
would require better localization precision than currently estimated. 

 The net result is
that between a few to a few tens of detections per year are expected by {\it
XEUS} with a follow-up of a short GRB. 
Assuming sufficiently good localization,
re-pointing after a gravitational signal detection can result in $\sim 4-54$ wind
detection per year from NS-NS mergers, for both {\it Con-X} and {\it XEUS}.
 Furthermore for {\it XEUS},
there is the exciting possibility to observe X-ray emission from BH-NS
mergers: $\sim 1-53$ event per year. The X-ray emission from these sources
should also be brighter than from a NS-NS mergers, since the mass of
the central BH could be much larger. 
 The above rates,
however, should be taken as indicative of upper limits. We
have not taken into account selection effects such as background/foreground 
sources and the fact that not all BH-NS and NS-NS mergers seem to
lead to an accreting system \citep[e.g.,][]{ross05,bel08}. Moreover, in some cases,
the X-ray afterglow from the burst could outshine the wind emission.
Nevertheless, the possibility to get information on mergers of compact
objects from electromagnetic signals remains, and it could bring
important understanding of the physics of these systems.

Finally, our findings have implications for interpreting late time activity observed in 
GRB afterglows.
We consider unlikely that fallback matter can be held responsible, since most of the mass is blown away.
Even if $\sim 10 \%$ of this matter can accrete all the way to the hole,
it is very unlikely that it could produce the observed flares, which have an energy ($\sim 10^{49}-10^{46}$ ergs) 
comparable to that of the prompt emission  \citep[e.g.][]{campana06}.
This would require that  the eccentric tail is far more massive than the main disc (contrary to what is observed in
simulations) or that the efficiency in converting accreted mass to energy 
 is somehow strongly enhanced in the late fallback accretion.
These arguments also apply  to the late accretion from the main disc, which is highly super-Eddington \citep{metzger08}.
We thus conclude that, in general, standard late time accretion is unlikely to account for 
the phenomena, like flares and plateaux, observed in GRB afterglows.

\section*{Acknowledgments}
The authors acknowledge useful discussions with K. Belczynski,  P. Bender, P. Armitage and  E. Quataert.
EMR acknowledges support from NASA though Chandra Postdoctoral Fellowship grant number 
PF5-60040 awarded by the Chandra X-ray Center, which is operated by the Smithsonian 
Astrophysical Observatory for NASA under contract NASA8-03060. MCB acknowledges support through
NASA Astrophysics Theory grant NNG06GI06G.

\end{document}